\documentclass[letterpaper,titlepage,11pt]{article}

\pdfoutput=1

\usepackage{amssymb,amsmath,amsfonts}
\usepackage{epsfig}
\usepackage{ccaption}
\usepackage{graphicx}
\usepackage{todonotes}
\usepackage[T1]{fontenc}

\usepackage[
      colorlinks=true,
      linkcolor=blue,
      urlcolor=magenta,
      filecolor=green,
      citecolor=red,
      pdfstartview=FitV,
      pdftitle={},
        pdfauthor={Stephane Detournay, Daniel Grumiller, Friedrich Scholler, Joan Simon},
        pdfsubject={},
        pdfkeywords={},
        pdfpagemode=None,
        bookmarksopen=true
      ]{hyperref}

\def\clock{{\count0=\time
           \divide\count0 60
           \ifnum\count0<10 0\fi\the\count0
           \multiply\count0 -60 \advance\count0 \time
           :\ifnum\count0<10 0\fi \the\count0
         }}
\newcommand{\timestamp}{{\small\vbox{\hbox{\tt\jobname.tex}
\hbox{\the\day/\the\month/\the\year, \clock}}}}

\usepackage{amsmath,amssymb}
\usepackage{verbatim}
\usepackage{graphicx}
\usepackage{hyperref}
\usepackage{color}

\DeclareFontFamily{OT1}{rsfs}{}
\DeclareFontShape{OT1}{rsfs}{m}{n}{ <-7> rsfs5 <7-10> rsfs7 <10->rsfs10}{} 
\DeclareMathAlphabet{\mycal}{OT1}{rsfs}{m}{n}

\newcommand{\be}[1]{ \begin{equation}\label{#1} }
\newcommand{\ee}{\end{equation}}
\newcommand{\bea}[1]{\begin{eqnarray}\label{#1} }
\newcommand{\eea}{\end{eqnarray}}
\newcommand{\p}{\partial}

\renewcommand{\phi}{\varphi}

\newcommand{\eq}[2]{\begin{equation} #1 \label{#2} \end{equation}}

\newcommand{\al}{\alpha}
\newcommand{\ga}{\gamma}
\newcommand{\de}{\delta}

\newcommand{\Om}{\Omega}

\DeclareMathOperator{\extdm}{d}
\newcommand{\extd}{\extdm \!}

\newcommand{\lin}{\psi} 

\begin{document}



\begin{titlepage}
\leftline{}{TUW--14--02}
\vskip 3cm
\centerline{\LARGE \bf Variational principle and 1-point functions}
\vskip 2mm
\centerline{\LARGE \bf in 3-dimensional flat space Einstein gravity} 
\vskip 1.6cm
\centerline{\bf Stephane Detournay$^{a}$, Daniel Grumiller$^{b}$, Friedrich Sch\"oller$^{b}$ and Joan Sim\'on$^{c}$}
\vskip 0.5cm
\centerline{\sl $^{a}$Physique Th\'eorique et Math\'ematique,
Universit\'e Libre de Bruxelles and International Solvay Institutes}
\centerline{\sl Campus Plaine C.P.~231, B-1050 Bruxelles, Belgium}
\smallskip
\centerline{\sl $^{b}$Institute for Theoretical Physics, Vienna University of Technology}
\centerline{\sl Wiedner Hauptstrasse 8-10/136, A-1040 Vienna, Austria}
\smallskip
\centerline{\sl $^{c}$School of Mathematics and Maxwell Institute for Mathematical Sciences}
\centerline{\sl University of Edinburgh, King's Buildings, Edinburgh EH9 3JZ, UK}
\vskip 0.5cm
\centerline{\small\tt sdetourn@ulb.ac.be,\, grumil@hep.itp.tuwien.ac.at,}
\centerline{\small\tt  schoeller@hep.itp.tuwien.ac.at, \, j.simon@ed.ac.uk}

\vskip 1.6cm
\centerline{\bf Abstract} \vskip 0.2cm \noindent

We provide a well-defined variational principle for 3-dimensional flat space Einstein gravity by adding one half of the Gibbons--Hawking--York boundary term to the bulk action. We check the 0-point function, recovering consistency with thermodynamics of flat space cosmologies. We then apply our result to calculate the 1-point functions in flat space Einstein gravity for the vacuum and all flat space cosmologies. The results are compatible with the ones for the zero mode charges obtained by canonical analysis.

\end{titlepage}
\pagestyle{empty}
\small
\normalsize
\newpage
\pagestyle{plain}
\setcounter{page}{1}


\addtocontents{toc}{\protect\setcounter{tocdepth}{2}}

\tableofcontents
\newpage

\section{Introduction}\label{se:1}

Given some bulk action $I[\Phi]$ and some boundary conditions on the fields $\Phi$, it is important to check whether or not there is a well-defined variational principle, $\delta I=0$ on-shell, for all variations $\delta\Phi$ that preserve the boundary conditions. Physically, the importance of $\delta I=0$ for solutions of the classical equations of motion (EOM) is evident: Only then solutions of the classical EOM are actually classical solutions, in the sense that they stabilize the action and allow for a meaningful (semi-)classical approximation to the path integral. 

Gravity theories are notorious in this regard, since the simplest way to resolve these issues is not accessible: Imposing natural boundary conditions on the metric, $g_{\mu\nu}\to 0$ would be most unnatural, as the metric must be non-degenerate near the boundary. For any given boundary value problem (for instance, a Dirichlet, Neumann or Robin boundary value problem) it is thus important to check whether or not there is a well-defined variational principle. 

If there is no well-defined variational principle, i.e., the first variation of the action does not vanish on-shell, then suitable boundary terms must be added to the bulk action to achieve a vanishing first variation.  A textbook example is provided by the pure Einstein-Hilbert action, whose variation yields a boundary term involving the metric variation, but also its first derivatives in the direction normal to the boundary. An additional term -- the Gibbons--Hawking--York term -- can then be added to the original action to compensate these unwanted terms and make the variational principle well-defined for variations such that $\delta g_{\mu \nu} = 0$ at the boundary (Dirichlet problem) (see \cite{waldgeneral}, App. E). Another archetypical situation is provided by AdS gravity, where after the addition of the Gibbons--Hawking--York term and appropriate counterterms, the Dirichlet variational principle is well-defined for variations $\delta g_{(0)ij}$ keeping the boundary metric fixed \cite{Papadimitriou:2005ii} (note that other variational problems can be posed for AdS$_3$ gravity, see \cite{Compere:2008us} for Neumann conditions and \cite{Compere:2013bya, Troessaert:2013fma} for mixed conditions). In general however, if  achieving a vanishing first variation turns out to be impossible, then the particular set of boundary conditions/boundary value problem must be discarded as unphysical.


Insisting on a well-defined variational principle is not only a matter of internal consistency, but has physical consequences. For example, the free energy as derived from the Euclidean path integral approach can only be correct if the action obeys a well-defined variational principle. Also holographic response functions are sensitive to the boundary terms that are added to the bulk action.

The main purpose of the present work is to establish a well-defined variational principle for flat space Einstein gravity in three dimensions and to calculate the 0- and 1-point functions, in particular for the flat space vacuum and for flat space cosmology solutions.

This paper is organized as follows. 
In section \ref{se:2} we review the situation for Einstein gravity with negative cosmological constant.
In section \ref{se:3} we discuss flat space Einstein gravity, provide a well-defined variational principle, check the free energy and calculate the holographic response functions.
In section \ref{se:4} we point to some loose ends in flat space holography.

Before starting we mention some of our conventions. We work exclusively in Euclidean signature. Our sign conventions for the Ricci tensor are fixed by $R_{\mu\nu}=\partial_\alpha\Gamma^\alpha{}_{\mu\nu}+\dots$.
We always consider 3-dimensional manifolds $M$ that are either topologically filled cylinders or filled tori. Their boundary is denoted by $\partial M$ and is then either topologically a cylinder or a torus.

\section{Einstein gravity with negative cosmological constant}\label{se:2}

Before studying the variational principle in 3-dimensional flat space, we revisit it in AdS$_3$ to understand whether there exists a smooth limit to the case of vanishing cosmological constant. None of the results in this section are new, but they provide a useful starting point for the novel results in section \ref{se:3}.

In section \ref{se:2.0} we recall the Brown--Henneaux boundary conditions.
In section \ref{se:2.1} we review the variational principle for various boundary value problems.
In section \ref{se:2.2} we recover the 0-point function and Ba\~nados--Teitelboim--Zanelli (BTZ) black hole thermodynamics.
In section \ref{se:2.3} we reconsider 1-point functions.

\subsection{Euclidean Anti-de~Sitter boundary conditions}\label{se:2.0}

Asymptotically AdS{}$_3$ metrics satisfying Brown--Henneaux boundary conditions are defined by \cite{Brown:1986nw}
\begin{equation}
\begin{aligned}
  g_{rr} &= \frac{\ell^2}{r^2} + h_{rr}(t,\,\varphi)\,\ell^4/r^4 + o(1/r^4) & g_{rt} &= {\cal O}(1/r^3) \\
  g_{tt} &= \frac{r^2}{\ell^2} + h_{tt}(t,\,\varphi) + o(1) & g_{r\phi} &= {\cal O}(1/r^3) \\
  g_{\phi\phi} &= r^2 + h_{\phi\phi}(t,\,\varphi)\,\ell^2 + o(1) & g_{t\phi} &= h_{t\phi}(t,\,\varphi)\,\ell + {\cal O}(1/r) \; .
\end{aligned}
\label{eq:one}
\end{equation}
The notation ${\cal O}(r^{-n})$ [$o(r^{-n})$] means that the quantity scales like $r^{-n}$ or smaller [smaller than  $r^{-n}$].
We consider the following set of variations preserving these boundary conditions (for convenience we scale all fluctuations $\delta h_{\mu\nu}$ with appropriate factors of the AdS radius $\ell$ so that they are dimensionless quantities)
\begin{equation}
\begin{aligned}
  \delta g_{rr} &= \delta h_{rr}(t,\phi) \,\ell^4/r^4 + o(1/r^4) & \delta g_{rt} &= {\cal O}(1/r^3) \\
  \delta g_{tt} &= \delta h_{tt}(t,\phi)  + o(1) &  \delta g_{r\phi} &= {\cal O}(1/r^3) \\
  \delta g_{\phi\phi} &= \delta h_{\phi\phi}(t,\phi)\,\ell^2 + o(1)  & \delta g_{t\phi} &= {\cal O}(1) \; .
\label{eq:vp4}
\end{aligned}
\end{equation}
While more general boundary conditions compatible with asymptotic AdS behavior are possible, for 3-dimensional Einstein gravity we do not need more general ones. All interesting physical solutions/states/fluctuations of Einstein gravity are already allowed by the boundary conditions above.

For later purposes we perform a $2+1$ split of the metric, $g_{ab}=\ga_{ab}+n_a n_b$, where $\ga_{ab}$ is the boundary metric and $n^a$ is the outward pointing unit normal vector. We always take an $r=\rm const.$ hypersurface as boundary. Extrinsic curvature is given by the standard expression $K^{ab}=\ga^{ac}\ga^{bd}\,\nabla_c n_d$. For a collection of useful formulas for variations and boundary quantities see \cite{McNees:Notes}.

The boundary conditions \eqref{eq:vp4} yield the following identities:
\begin{subequations}
 \label{eq:vp13}
\begin{align}
  n_a &= \de_a^r\,\frac{\ell}{r} + {\cal O}(1/r^3) \\
  \sqrt{\gamma} &= \frac{r^2}{\ell} + {\cal O}(1) \\
  K &= \frac{2}{\ell} - \big(h_{tt}+h_{\phi\phi}+h_{rr}\big)\,\frac{\ell}{r^2} + {\cal O}(1/r^4)  \label{eq:vp666} \\
  \gamma^{ab} \delta g_{ab} &= (\delta h_{\phi\phi} + \delta h_{tt})\,\frac{\ell^2}{r^2} + o(1/r^2) \\
  K^{ab} \delta g_{ab} &= (\delta h_{\phi\phi} +  \delta h_{tt})\,\frac{\ell}{r^2} + o(1/r^2) \\
  n^an^b \delta g_{ab} &= \delta h_{rr}\,\frac{\ell^2}{r^2} + o(1/r^2) \\
  \gamma^{ab} n^c \nabla_c \delta g_{ab} &= - 2 (\delta h_{\phi\phi} +  \delta h_{tt})\,\frac{\ell}{r^2} + o(1/r^2) 
\end{align}
\end{subequations}

\subsection{Variational principle}\label{se:2.1}

We consider the Einstein--Hilbert (EH) action including a cosmological constant with a Gibbons--Hawking--York (GHY) \cite{Gibbons:1976ue,York:1972sj} and a cosmological boundary term with arbitrary dimensionless coefficients $\alpha$ and $\beta$, respectively,
\begin{equation}
  \Gamma_{(\alpha,\, \beta)} = - \frac{1}{16\pi G}\, \int_M\!\!\! \mathrm{d}^3x \sqrt{g}\, \Big( R - 2 \Lambda\Big) - \frac{1}{8\pi G}\, \int_{\partial M}\!\!\! \!\!\!  \mathrm{d}^2x \sqrt{\gamma}\, \Big( \alpha K + \frac\beta\ell \Big) \,,
\label{eq:ads}
\end{equation}
where $\Lambda = - \tfrac{1}{\ell^2}$. The first variation of the bulk action \eqref{eq:ads} equals 
\begin{align}
  \delta\Gamma_{(\alpha,\, \beta)}
    &= \frac{1}{16\pi G}\, \int_M\!\!\! \mathrm{d}^3x \sqrt{g}\, \Big( G^{ab} + \Lambda g^{ab} \Big) \,\delta g_{ab}  \nonumber \\
    &\quad + \frac{1}{16\pi G}\, \int_{\partial M}\!\!\!\!\!\! \mathrm{d}^2x \sqrt{\gamma} \,\Big( K^{ab} - \big(\alpha K + \frac\beta\ell\big) \gamma^{ab} + \big(\alpha - 1\big) K n^a n^b \Big) \,\delta g_{ab} \nonumber  \\
    &\quad + \frac{1 - \alpha}{16\pi G}\, \int_{\partial M}\!\!\!\!\!\! \mathrm{d}^2x \sqrt{\gamma} \,\gamma^{ab} n^c \nabla_c \, \delta g_{ab}  \nonumber \\
    &\quad + \frac{2\alpha - 1}{16\pi G} \, \int_{\partial^2 M}\!\!\!\!\!\!\!\! \mathrm{d}x \sqrt{\gamma'}\, n'^a n^b \, \delta g_{ab} \; .  \label{eq:adsvar}
\end{align}
Primed quantities are corner contributions that arise in case of non-smooth boundaries, see e.g.~\cite{Farhi:1989yr,Hayward:1993my,Brill:1994mb,Fursaev:1995ef,Hawking:1996ww}. We assume for the remainder of our AdS${}_3$ review that there are no corner terms.

The identities \eqref{eq:vp13} allow to write the variation $\delta\Gamma_{(\alpha,\, \beta)}$ as
\begin{equation}
  \delta\Gamma_{(\alpha,\, \beta)} =
  \frac{1}{16\pi G}\, \int_{\partial M}\!\!\!\!\!\! \mathrm{d}^2x \, \big(
    2 (\alpha-1)\, \delta h_{rr}
    - (\beta+1)\,(\delta h_{tt} + \delta h_{\phi\phi}) \big) 
      + o(1) \; .
 \label{eq:ons-var}
\end{equation}
Since we are interested in a well-defined variational principle, we study the situations under which this variation vanishes. 

The first observation is that if we choose
\begin{equation}
  \alpha = 1 \qquad \beta = - 1
\label{eq:angelinajolie}
\end{equation}
no matter whether the variations $\delta h_{ab}$ are on-shell or not, the first variation \eqref{eq:ons-var} vanishes, leading to a well defined variational principle for the well known action \cite{Henningson:1998gx,
Balasubramanian:1999re,Emparan:1999pm}
\begin{equation}
  \Gamma = - \frac{1}{16\pi G}\, \int_M \!\!\!\mathrm{d}^3x \sqrt{g}\, \Big( R + \frac{2}{\ell^2} \Big)
  - \frac{1}{8\pi G}\, \int_{\partial M}\!\!\!\!\!\! \mathrm{d}^2x \sqrt{\gamma}\, \Big( K - \frac{1}{\ell} \Big) \; .
\end{equation}

To explore whether there exist more general situations, we study the linearised EOM satisfied by the variations $\de h_{ab} = g_{ab}-\bar{g}_{ab}$, where $\bar{g}_{ab}$ stands for the AdS${}_3$ metric. These are given by \cite{Deser:2003vh}
\begin{align}
  G^{(1)}_{\mu\nu} &= R_{\mu\nu}^{(1)} - \frac{1}{2}\bar{g}_{\mu\nu}\,R^{(1)} - 2\Lambda\,\de h_{\mu\nu} = 0 \\
  R_{\mu\nu}^{(1)} &= \frac{1}{2}\big(-\bar{\nabla}^2 \,\de h_{\mu\nu} - \bar{\nabla}_\mu\bar{\nabla}_\nu\,\de h + \bar{\nabla}^\sigma\bar{\nabla}_\nu \,\de h_{\sigma\mu} + \bar{\nabla}^\sigma\bar\nabla_\mu \,\de h_{\sigma\nu}\big) \\
  R^{(1)} &= -\bar\nabla^2 \,\de h + \bar\nabla_\mu\bar\nabla_\nu \,\de h^{\mu\nu} - 2\Lambda\,\de h
\end{align}
where all barred quantities are computed using the AdS${}_3$ metric. If we take the trace of the linearised Einstein equation, we learn that $R^{(1)}=0$. 

In transverse gauge for the fluctuations, $\nabla^\mu (\delta h_{\mu\nu} - g_{\mu\nu}g^{\rho\sigma}\delta h_{\rho\sigma})=0$, the linearised EOM imply tracelessness (see e.g.~\cite{Li:2008dq}).
\begin{equation}
  \delta h_{tt} + \delta h_{\phi\phi} + \delta h_{rr} = 0
 \label{eq:tracevan}
\end{equation}
Inserting this asymptotic on-shell constraint into the variation \eqref{eq:ons-var}, one obtains
\begin{equation}
  \left.\delta\Gamma_{(\alpha,\, \beta)}\right|_{\mathrm{EOM}} =
  \frac{1}{16\pi G}\, \int_{\partial M} \!\!\!\!\!\! \mathrm{d}^2x\, \delta h_{rr}\, \big(2\alpha + \beta - 1\big) + o(1)\,.
\end{equation}
Thus, for variations preserving the boundary conditions \eqref{eq:vp4} and tracelessness \eqref{eq:tracevan}, there is a 1-parameter family of actions \eqref{eq:ads} with a well-defined variational principle, determined by the condition
\begin{equation}
  2\alpha = 1 - \beta\,.
\label{eq:vp1}
\end{equation}
This result is expected on general grounds, see e.g.~\cite{Papadimitriou:2004ap}. Their key observation is that asymptotically there is no difference between an extrinsic curvature term and a boundary cosmological constant, see \eqref{eq:vp666}. Thus only the value of $2\alpha+\beta$ is of relevance, this being fixed to unity by the condition \eqref{eq:vp1}.\footnote{%
DG thanks Ioannis Papadimitriou for explanations concerning this issue.
} 

Note that it is rather unusual to use the linearized equations of motion for the fluctuations --- here, relation \eqref{eq:tracevan} --- in order for the variational principle to be well-defined. Indeed, one would normally expect to have $\delta \Gamma = 0$ {\itshape for all $\delta g_{\mu \nu}$ allowed by the boundary conditions} defining the theory. In the AdS$_3$ case considered here, the tracelessness condition \eqref{eq:tracevan} could be interpreted as restricting the Brown--Henneaux boundary conditions \eqref{eq:vp4}. One might worry whether this is too restrictive to yield a sensible theory. However, the Virasoro descendants of the BTZ black holes satisfy the tracelessness condition. Also, one can check that this condition is invariant under the action of the Brown--Henneaux generators. Therefore, this restricted set of boundary conditions not only contains the physically most relevant solutions but still displays the two Virasoro algebras characteristic of two-dimensional conformal symmetry.

Besides the special choice \eqref{eq:angelinajolie} that leads to the standard situation of a Dirichlet problem for the metric, a GHY term with the usual normalization and the well-known holographic counterterm, there is another case that stands out. Namely, for the choice \footnote{%
A similar choice was studied in arbitrary dimensions in \cite{Mora:2004kb}, see also \cite{Mora:2006ka,Miskovic:2006tm}. DG thanks Alfredo Perez, David Tempo and Ricardo Troncoso for several discussions on this topic.
}
\eq{
\alpha=\frac12\qquad \beta = 0
}{eq:lalapetz}
the boundary action is independent from the cosmological constant [and the corner term in \eqref{eq:adsvar} vanishes]. In this case there is no holographic counterterm, and the GHY term does not have its usual normalization, but has an additional factor of $\tfrac12$. The independence from the cosmological constant makes this choice most suitable for an $\ell\to\infty$ limit.

\subsection{0-point function}\label{se:2.2}

Evaluating the action \eqref{eq:ads} with the condition \eqref{eq:vp1} on classical solutions leads to the free energy, $F=T\, \Gamma_{(\alpha,\, \beta)}$, where $T$ is the temperature, determined by the inverse of the periodicity of Euclidean time. 
\eq{
F = \lim_{r_c\to\infty}\,\frac{1}{4G}\,\Big(\frac{r_c^2-r_+^2}{\ell^2}-(2\al+\beta)\frac{r_c}{\ell}\,\sqrt{g_{tt}}\Big)
}{eq:vp9}
The first term comes from the bulk action, integrated from the center or outer horizon, $r=r_+$, to the asymptotic boundary, which we regulate by a cut-off $r=r_c$ that we send to infinity at the end of the calculation. The second term comes from the boundary action evaluated at the same value of the cut-off. Note that the result for the free energy is universal and finite as long as $\al$ and $\beta$ satisfy the condition \eqref{eq:vp1} and the metric obeys the boundary conditions \eqref{eq:one}.

For instance, taking BTZ black holes \cite{Banados:1992wn,Banados:1992gq}
\begin{equation}
  \mathrm{d}s^2 = -\frac{(r^2 - r_+^2)(r^2 - r_-^2)}{\ell^2r^2}\, \mathrm{d}t^2
  + \frac{\ell^2r^2}{(r^2 - r_+^2)(r^2 - r_-^2)} \,\mathrm{d}r^2 + r^2 \,\big( \mathrm{d}\phi - \frac{r_+r_-}{\ell r^2} \,\mathrm{d}t \big)^2
\label{eq:BTZ}
\end{equation}
and inserting their Euclidean continuation into the formulas above leads to the free energy [$T=(r_+^2-r_-^2)/(2\pi r_+\ell^2)$, $\Om=r_-/(r_+\ell)$]
\eq{
F(T,\ \Omega) = -\frac{r_+^2-r_-^2}{8G\ell^2} = -\frac{\pi^2 T^2 \ell^2}{2G\,(1-\Omega^2\ell^2)}
}{eq:vp2}
independently from $\alpha$ and $\beta$, provided they obey the relation \eqref{eq:vp1}.

The result \eqref{eq:vp2} leads to the correct mass, $M=(r_+^2+r_-^2)/(8G\ell^2)$, and entropy, $S=2\pi r_+/(4G)$, of BTZ black holes, and is consistent with the first law of thermodynamics, $\extd F=-S\extd T-J\extd \Omega$, where $J=r_+r_-/(4G\ell)$. We stress that this would no longer be the case if $\alpha$ and $\beta$ violate the condition \eqref{eq:vp1}. Thus, the BTZ free energy emerges correctly if and only if we have a well-defined variational principle. 

\subsection{1-point functions}\label{se:2.3}


We focus first on the standard case $\alpha=1$, $\beta=-1$.
To compute the 1-point functions, we consider the same asymptotic metric as in \eqref{eq:one}, but we allow a different set of metric fluctuations to accommodate for sources
\begin{equation}
\begin{aligned}
  \delta g_{rr} &= \text{irrelevant} & \delta g_{rt} &= \text{irrelevant} \\
  \delta g_{tt} &= \delta h_{tt}^{(0)}(t,\phi) \,r^2/\ell^2 + \delta h_{tt}(t,\phi) + o(1) & \delta g_{\phi\phi} &= \delta h_{\phi\phi}^{(0)}(t,\phi) \,r^2 + \delta h_{\phi\phi}(t,\phi)\,\ell^2 + o(1) \\
  \delta g_{r\phi} &= \text{irrelevant} & \delta g_{t\phi} &= \delta h_{t\phi}^{(0)}(t,\phi) \,r^2/\ell + {\cal O}(1) \;.
\label{eq:vp3}
\end{aligned}
\end{equation}
Notice that these fluctuations do not respect the Brown--Henneaux boundary conditions \eqref{eq:vp4} in general. The leading contribution, in an ${\cal O}(1/r)$ expansion, of the on-shell variation of the classical action equals
\begin{equation}
\begin{aligned}
  \left.\delta\Gamma\right|_{\mathrm{EOM}} &=
  \frac{1}{16\pi G}\, \int_{\partial M}\!\!\!\!\!\! \mathrm{d}^2x \,\Big(
    \big( \frac{h_{rr}}{2} + h_{\phi\phi} \big)\, \delta h_{tt}^{(0)}
    - 2 h_{t\phi} \,\delta h_{t\phi}^{(0)}
    + \big( \frac{h_{rr}}{2} + h_{tt} \big)\, \delta h_{\phi\phi}^{(0)}
  \Big) 
\end{aligned}
\label{eq:source-var}
\end{equation}
%
The fluctuations \eqref{eq:vp3} are non-normalizable and are interpreted as sources for the corresponding operators in the dual CFT \cite{Witten:1998qj}. 
 
Let us evaluate this variation for the (Euclidean version of the) BTZ black hole \eqref{eq:BTZ}.
The non-trivial functions $h_{ab}$ are identified to be
\begin{align}
   h_{rr} = - h_{tt} &= \frac{r_+^2 + r_-^2}{\ell^2} & h_{t\phi} &= i\, \frac{r_+ r_-}{\ell^2}\,,
\label{eq:vp42}
\end{align}
giving rise to an on-shell variation \eqref{eq:source-var}
\begin{equation}
\begin{aligned}
  \left.\delta\Gamma\right|_{\mathrm{EOM}} &=
  \frac{1}{32\pi G\,\ell^2} \, \int_{\partial M} \!\!\!\!\!\!\mathrm{d}^2x \,\big(
    (r_+^2 + r_-^2) \, (\delta h_{tt}^{(0)} - \delta h_{\phi\phi}^{(0)}) - 4i r_+ r_- \,\delta h_{t\phi}^{(0)}
  \big) \;. 
\label{eq:whatever}
\end{aligned}
\end{equation}
This is the usual ``holographic'' structure \cite{Witten:1998qj} of $\delta\Gamma|_{\mathrm{EOM}} \sim \textrm{(vev)}\,\delta\textrm{(source)}$, where `vev' here refers to the holographically renormalized Brown--York stress tensor, $T^{ab}$, 
and `source' to the non-normalizable fluctuations, $h_{ab}^{(0)}$, that we switched on in order to get a finite response. Evaluating the holographically renormalized Brown--York stress tensor
\eq{
\delta\Gamma|_{\mathrm{EOM}} = -\frac12\,\int\extd^2x\sqrt{g^{(0)}}\,T^{ab}\,\de h_{ab}^{(0)}
}{eq:BY}
on the (Euclidean version of) BTZ solutions \eqref{eq:BTZ} using our result \eqref{eq:whatever} yields (here we set $\ell=1$ and provide only absolute values for easier comparison with Minkowskian results)
\eq{
|T_{tt}|=|T_{\phi\phi}| = \frac{r_+^2+r_-^2}{16\pi G} = \frac{M}{2\pi} \qquad \qquad |T_{t\phi}| = \frac{|r_+r_-|}{8\pi G} = \frac{J}{2\pi}\;.
}{eq:BYresult}
The expressions \eqref{eq:BYresult} are well-known and reproduce precisely previous calculations, see e.g.~\cite{Balasubramanian:1999re,Kraus:2005zm} and references therein.

For other values of $\alpha$ and $\beta$, subject to the condition \eqref{eq:vp1}, we obtain the same response functions, as expected. 

\section{Flat space Einstein gravity}\label{se:3}

Equipped with the explicit results for Euclidean AdS$_3$ from the previous section, we address in this section the situation in flat space Einstein gravity. 

In section \ref{se:3.0} we formulate a specific set of flat space  boundary conditions.
In section \ref{se:3.1} we provide the variational principle for Einstein gravity compatible with these boundary conditions. 
In section \ref{se:3.2} we check the 0-point function and flat space cosmology thermodynamics.
In section \ref{se:3.3} we calculate 1-point functions.

\subsection{Euclidean flat space boundary conditions}\label{se:3.0}

The flat space boundary conditions are usually provided in Eddington--Finkelstein (EF) gauge adapted to null infinity \cite{Barnich:2006av}.\footnote{There is earlier work where asymptotic flatness was defined in space polar coordinates and time, for example see 
\cite{Ashtekar:1993ds,Marolf:2006xj}.} However, for Euclidean purposes this gauge is inaccessible since there is no Euclidean analog of a null vector. Thus, our first task is to translate these boundary conditions into a more suitable gauge.

We start with the set of boundary conditions provided in \cite{Bagchi:2012yk} (these are looser boundary conditions than the ones by Barnich and Comp\`ere \cite{Barnich:2006av}).\footnote{%
In particular, the function $h_3$ does not appear in \cite{Barnich:2006av}, so the central charge term does not emerge from \eqref{eq:vp41}, but instead from the (modified) transformation behavior of $h_{uu}$. Their boundary conditions are preserved by Lie variations along vector fields \eqref{eq:vp38}, \eqref{eq:vp39}, provided a specific trivial contribution \eqref{eq:vp40} with $f_1=\xi_L'''$ is added to the generators. We thank Geoffrey Comp\`ere for explaining why their boundary conditions are consistent.
}
\begin{equation}
\begin{aligned}
  g_{uu} &= h_{uu} + {\cal O}(1/r) & g_{u\phi} &= h_{u\phi} + {\cal O}(1/r) \\
  g_{ur} &= -1 + h_{ur}/r + {\cal O}(1/r^2)  & g_{r\phi} &= h_1 (\phi)  + {\cal O}(1/r) \\
  g_{\phi\phi} &= r^2 + \big(h_2(\phi) + u h_3(\phi)\big) r + {\cal O}(1)  & g_{rr} &= {\cal O}(r^{-2}) \;.
\label{eq:vp14}
\end{aligned}
\end{equation}
The Minkowski background is given by $h_{uu}=-1$ with the remaining $h_{\mu\nu}=h_i=0$ and all subleading terms set to zero, $\extd s^2=-\extd u^2-2\extd u\extd r+r^2\,\extd\phi^2$.

The asymptotic symmetry group is generated by vector fields
\begin{align}
 \xi_L &= \xi_L(\phi)\, \p_\phi + \xi^\prime_L(\phi)\, \big(u \p_u - r\p_r\big) - \xi''_L(\phi) \,\frac{u}{r}\,\p_\phi + \dots \label{eq:vp38}\\ 
 \xi_M &= \xi_M(\phi)\, \p_u +\label{eq:vp39}\\
 &+ {\cal O}(1/r) \p_u + \big(u f_1(\phi) + f_2(\phi) + {\cal O}(1/r)\big)\p_r + \big(f_3(\phi)/r + {\cal O}(1/r^2)\big) \p_\phi \label{eq:vp40}
\end{align}
where the dots refer to sub-leading terms generating trivial gauge transformations which are modded out in the asymptotic symmetry group. 

The equations of motion impose additional relations between the free functions appearing in \eqref{eq:vp14}:
\begin{align}
 \partial_u h_{uu} &= 0 \\
 \partial_u h_{rr} &= -2 h_{ur} \\
 \partial_u h_{r\phi} &= u\partial_\phi h_{uu} + \partial_\phi h_{ur} - 2 h_{u\phi} + h_4(\phi)
\end{align}
Note that the integration function $h_4(\phi)$ erroneously was set to zero in the equation displayed in the text below Eq.~(7b) in \cite{Bagchi:2012yk}. While this omission is not relevant for flat space chiral gravity, it is important to take into account this function in Einstein gravity.

It is of interest to derive the centrally extended BMS$_3$ algebra \cite{Bondi:1962,Sachs:1962,Barnich:2006av} as the asymptotic symmetry algebra due to the boundary conditions \eqref{eq:vp14}. Since the canonical charges $Q_M$ associated with the super-translation generators $M$ \cite{Bagchi:2012yk} are $Q_M \sim \int d\varphi \, \xi_M(\varphi) (h_{uu} + h_3)$, we can answer this question by computing
the variation of the state dependent function $h_{uu}+h_3$. Using the on-shell relation $\partial_u h_{uu}=0$, we find from the Lie-derivatives along $\xi_{L,\,M}$
\begin{align}
 \delta_{\xi_M} h_{uu} &= 0 +\dots & \delta_{\xi_L} h_{uu} &= h_{uu}^\prime\xi_L + 2h_{uu}\xi_L^\prime +\dots\\
 \delta_{\xi_M} h_3 &= 0 +\dots & \delta_{\xi_L} h_3 &= h_3^\prime\xi_L + 2h_3\xi_L^\prime - 2 \xi_L''' +\dots
\label{eq:vp41}
\end{align}
From the left hand equations we can read off the vanishing commutator between the super-translation generators, while the right hand equations yield the expected Schwarzian derivative for the combination $h_{uu}+h_3$, including the anomalous term. With the appropriate normalization of the charges $Q_M$, the correct central charge $c_M=3/G$ is reproduced \cite{Barnich:2006av} and the asymptotic symmetry algebra matches the centrally extended BMS$_3$ algebra, which is isomorphic to a 2-dimensional Galilean conformal algebra \cite{Bagchi:2010zz}.

We convert now the boundary conditions from EF gauge \eqref{eq:vp14} into diagonal gauge, starting from the line-element
\begin{equation}
  \extd s^2 = -f(\phi)\,\extd u^2 - 2\extd u\,\extd r + r^2\,\extd\phi^2 + 2g(\phi)\,\extd u\,\extd\phi + h_1(\phi)\,\extd r\,\extd \phi + \dots
\end{equation}
where the ellipsis denotes subleading terms.
We define diagonal gauge as one in which there is no $\extd r\,\extd t$ term, where $t$ is the time coordinate replacing the retarded time $u$. 
\begin{equation}
  u = t + K(r,\phi) \quad \Rightarrow \qquad \extd u = \extd t + \partial_r K\,\extd r + \partial_\phi K\,\extd\phi\,.
\end{equation}
Absence of $\extd r\,\extd t$ terms yields the partial differential equation
\begin{equation}
  \partial_r K = -\frac{1}{f} \quad \Rightarrow \quad K(r,\phi) = -\frac{r}{f(\phi)} + K_0(\phi)\,.
\end{equation}
The transformed line-element is then
\begin{equation}
  \extd s^2 = -f(\phi)\,\extd t^2 + \frac{\extd r^2}{f(\phi)} + A\,\extd\phi^2 + 2B\,\extd t\,\extd\phi + 2C\,\extd r\,\extd\phi + \dots
\end{equation}
with the functions
\begin{align}
  A &= r^2 - f\left(\partial_\phi K\right)^2 + 2g\partial_\phi K
  = r^2\Big(1 - \frac{(f^\prime)^2}{f^3}\Big) +2r\,\frac{f^\prime}{f^2}\left(g- f K_0^\prime\right) + K_0^\prime\left(2g-fK_0^
  \prime\right) \\
  B &= -f\partial_\phi K + g = -r\frac{f^\prime}{f} - fK_0^\prime + g \\
  C &= -f\partial_\phi K \partial_r K - \partial_\phi K + g\partial_r K = h_1 - \frac{g}{f}\,.
\end{align}
From the expression for the function $A$ we see that only zero-mode solutions, $f'=0$, have the usual $r^2\,\extd\phi^2$ term in diagonal gauge. Restricting to constant $f$ also eliminates the term linear in $r$ in the function $B$. Since it is not obvious to us how to interpret the asymptotic line-element for non-constant $f$ we restrict from now on to constant $f$ and postpone comments on the general case to section \ref{se:4}. 

Converting the time $t$ to Euclidean time $\tau$, we shall study background metrics of the form
\begin{equation}
  \mathrm{d}s^2 =
   h_{\tau\tau}(\phi)\, \mathrm{d}\tau^2 + h_{rr}(\phi)\, \mathrm{d}r^2 +  r^2 \,\mathrm{d}\phi^2 \; ,
\label{eq:vp5}
\end{equation}
with $h_{rr}=1/h_{\tau\tau}$ and fluctuations satisfying 
\begin{equation}
\begin{aligned}
  \delta g_{\phi\phi} = {\cal O}(r) \qquad \delta g_{\phi\tau} &= {\cal O}(1) \qquad
  \delta g_{\tau\tau} \sim \delta g_{rr}  = {\cal O}(1) \\
  \delta g_{r\phi} = {\cal O}(1) \qquad \delta g_{r\tau} &= {\cal O}(1/r) \qquad
  \delta(g_{rr} g_{\tau\tau}) = {\cal O}(1/r) 
\end{aligned}
\label{eq:bcdiagonal}
\end{equation}
where to highest order $\delta g_{rr}$, $\delta g_{r\tau}$, and $\delta g_{\tau\tau}$ only depend on $\phi$. We stress that our subset of fluctuations $\delta g_{\tau\tau},\,\delta g_{rr}$ in \eqref{eq:bcdiagonal} differs from the ones considered in \cite{Ashtekar:1993ds,Marolf:2006xj}. In that respect, the theories studied in these references are different from ours. The conditions on the fluctuations $\delta g_{rr}$ and $\delta g_{\tau\tau}$ imply that we can parametrize them as
\eq{
\delta g_{\tau\tau} = \delta h_{\tau\tau} + {\cal O}(1/r) \qquad \delta g_{rr} = \delta h_{rr} + {\cal O}(1/r) \qquad \delta (h_{rr} h_{\tau\tau}) = 0 \,.
}{eq:vp8}

In summary, the boundary conditions above are a diagonal gauge version of the flat space boundary conditions imposed in Einstein gravity \cite{Barnich:2006av} or flat space chiral gravity \cite{Bagchi:2012yk}. They can be transformed into each other for zero-mode solutions [meaning that the $\phi$-dependent functions in \eqref{eq:vp5} actually are constant] using the standard coordinate transformation between EF and Schwarzschild coordinates. We do not address non-zero mode solutions in this work.\footnote{We stress that this restriction already covers physically interesting configurations as conical defects and flat cosmologies.}

For later purposes we collect here identities analog to \eqref{eq:vp13} for an $r=\rm const.$ boundary.
\begin{subequations}
 \label{eq:vp15}
\begin{align}
  n_a &= \de_a^r\,\sqrt{h_{rr}} + {\cal O}(1/r) \\
  \sqrt{\gamma} &= r \sqrt{h_{\tau\tau}} + {\cal O}(1) \\
  K &= \frac{1}{r \sqrt{h_{rr}}} + {\cal O}(1/r^2)
   \\
   g^{ab} \delta g_{ab} &= \frac{\delta h_{rr}}{h_{rr}} + \frac{\delta h_{\tau\tau}}{h_{\tau\tau}} + {\cal O}(1/r) \\
   K^{ab} \delta g_{ab} &= {\cal O}(1/r^2) \\
   \gamma^{ab} n^c \nabla_c \delta g_{ab} &= {\cal O}(1/r^2)
\end{align}
\end{subequations}

\subsection{Variational principle}\label{se:3.1}

We establish now that the EH action with the addition of one half of the GHY boundary term 
\begin{equation}
  \Gamma = - \frac{1}{16\pi G}\, \int_M\!\!\! \mathrm{d}^3x \sqrt{g}\, R - \frac{1}{16\pi G}\, \int_{\partial M} \!\!\!\!\!\!\mathrm{d}^2x \sqrt{\gamma}\, K \; ,
 \label{eq:3daction}
\end{equation}
gives rise to a well defined variational principle for flat space boundary conditions \eqref{eq:vp5}, \eqref{eq:bcdiagonal}. 

To prove this statement, we consider a one-parameter family of actions given by
\begin{align}
  \Gamma_{(\alpha)} &= - \frac{1}{16\pi G}\, \int_M \!\!\!\mathrm{d}^3x \sqrt{g}\, R - \frac{1}{8\pi G}\, \int_{\partial M} \!\!\!\!\!\!\mathrm{d}^2x \sqrt{\gamma}\,\alpha K \; .
\label{eq:vp6}
\end{align}
Note that \eqref{eq:vp6} is the $\ell\to\infty$ limit of the action \eqref{eq:ads}. Its one free parameter $\alpha$ will be determined by requiring the first variation
\begin{align}
  \delta\Gamma_{(\alpha)}
    &= \frac{1}{16\pi G}\, \int_M \!\!\!\mathrm{d}^3x \sqrt{g}\, G^{ab}\, \delta g_{ab} \nonumber  \\
    &\quad + \frac{1}{16\pi G}\, \int_{\partial M}\!\!\!\!\!\! \mathrm{d}^2x \sqrt{\gamma}\, \big( K^{ab} - \alpha K g^{ab} + (2\alpha - 1) K n^a n^b \big)\, \delta g_{ab} \nonumber \\
    &\quad + \frac{1 - \alpha}{16\pi G}\, \int_{\partial M}\!\!\!\!\!\! \mathrm{d}^2x \sqrt{\gamma}\, \gamma^{ab} n^c \nabla_c \, \delta g_{ab} \nonumber \\
    &\quad + \frac{2\alpha - 1}{16\pi G}\, \int_{\partial^2 M}\!\!\!\!\!\!\!\! \mathrm{d}x \sqrt{\gamma'}\, n'^a n^b \delta g_{ab} 
\label{eq:var1}
\end{align}
to vanish on-shell.  Notice we used $g^{ab}=\ga^{ab}+n^a n^b$ when writing the second line in \eqref{eq:var1}.

The identities \eqref{eq:vp15} allow to prove the relations
\begin{align}
&  \int_{\partial M} \!\!\!\!\!\!\mathrm{d}^2x \sqrt{\gamma}\, K^{ab} \,\delta g_{ab} \sim {\cal O}(1/r) \\
&  \int_{\partial M} \!\!\!\!\!\!\mathrm{d}^2x \sqrt{\gamma}\, K g^{ab} \,\delta g_{ab} = \int_{\partial M} \!\!\!\!\!\!\mathrm{d}^2x \,\sqrt{\frac{h_{\tau\tau}}{h_{rr}}} \Big( \frac{\delta h_{rr}}{h_{rr}} + \frac{\delta h_{\tau\tau}}{h_{\tau\tau}} \Big) + {\cal O}(1/r) \\
& \int_{\partial M} \!\!\!\!\!\!\mathrm{d}^2x \sqrt{\gamma}\, K n^a n^b \delta g_{ab} = \int_{\partial M} \!\!\!\!\!\!\mathrm{d}^2x \, \sqrt{\frac{h_{\tau\tau}}{h_{rr}^3}} \, \delta h_{rr} \\
&  \int_{\partial M} \!\!\!\!\!\!\mathrm{d}^2x \sqrt{\gamma}\, \gamma^{ab} n^c \nabla_c \, \delta g_{ab} \sim {\cal O}(1/r)\;.
\end{align}
Thus, the first variation of the action \eqref{eq:3daction} yields on-shell
\begin{equation}
\begin{aligned}
  \left.\delta\Gamma_{(\alpha)}\right|_{\mathrm{EOM}} =& - \frac{\alpha}{16\pi G}\, \int_{\partial M}\!\!\!\!\!\! \mathrm{d}^2x \,\sqrt{\frac{h_{\tau\tau}}{h_{rr}}} \, \frac{\delta(h_{rr}h_{\tau\tau})}{h_{rr}h_{\tau\tau}} + {\cal O}(1/r) \\
  &+ \frac{2\alpha -1}{16\pi G}\, \Big( \int_{\partial M}\!\!\!\!\!\! \mathrm{d}^2x \, \sqrt{\frac{h_{\tau\tau}}{h_{rr}^3}} \, \delta h_{rr} + \int_{\partial^2 M}\!\!\!\!\!\!\!\! \mathrm{d}x \sqrt{\gamma'}\, n'^a n^b \delta g_{ab} \Big)  \; .
\label{eq:vp7}
\end{aligned}
\end{equation}
The first line in \eqref{eq:vp7} vanishes, for any value of $\alpha$, as $r$ goes to infinity due to \eqref{eq:vp8}. The point we want to stress is that even in the absence of corner terms, the second line in \eqref{eq:vp7} does not vanish for all fluctuations $\delta h_{rr}$ in  \eqref{eq:vp8}.\footnote{Notice that for the boundary conditions considered in \cite{Marolf:2006xj}, this last term would also be subleading, since those boundary conditions impose stronger fall-off behavior on the fluctuations than ours. Thus, we reproduce the claim in \cite{Marolf:2006xj} that there exists a well defined variational principle for $\alpha=1$. In fact, we showed above that this claim is true for {\it any} $\alpha$.} Thus, requiring the existence of a well defined variational principle for the set of boundary conditions \eqref{eq:bcdiagonal} fixes the value of $\alpha = \frac{1}{2}$. Notice that this value would also ensure the vanishing of the corner terms, if these were to exist. Thus,
\begin{equation}
  \left.\delta\Gamma_{(\frac{1}{2})}\right|_{\mathrm{EOM}} = {\cal O}(1/r)\,.
\end{equation}
The standard GHY boundary term corresponds to the choice $\alpha=1$. Our analysis in \eqref{eq:vp7} proves the lack of a well defined variational principle for this boundary term for the flat space boundary conditions \eqref{eq:vp5}, \eqref{eq:bcdiagonal}. It is interesting to point out that our answer $\alpha=\tfrac12$ is compatible with the consistency relation \eqref{eq:vp1} together with the fact that in flat space $\beta$ is effectively zero.

In conclusion, the action \eqref{eq:3daction}, which contains one half of the usual GHY boundary term, with flat space boundary conditions \eqref{eq:vp5}, \eqref{eq:bcdiagonal} leads to a well-defined variational principle. The action \eqref{eq:3daction} arises as a smooth $\ell\to\infty$ limit of the AdS action \eqref{eq:ads} with the choice \eqref{eq:lalapetz}.

\subsection{0-point function}\label{se:3.2}

Like in the AdS case, the action \eqref{eq:3daction} can be used to determine the free energy of specific classical solutions. 
\eq{
F = \lim_{r_c\to\infty} -\frac{\alpha}{4 G}\,\sqrt{\gamma}\, K\, \Big|_{r=r_c} = -\frac{\alpha}{4 G}\,\sqrt{\frac{h_{\tau\tau}}{h_{rr}}}\,\Big|_{r\to\infty}
}{eq:vp12}
Our result for free energy is finite, but does depend on the choice of $\alpha$.

The solutions of interest here are the flat vacuum and flat space cosmologies \cite{Cornalba:2002fi,Cornalba:2002nv,Cornalba:2003kd}. Their Euclidean version is given by the line-element \cite{Bagchi:2013lma}
\eq{
\extd s^2 = r_+^2\,\Big(1-\tfrac{r_0^2}{r^2}\Big)\,\extd\tau^2 + \frac{\extd r^2}{r_+^2\,\big(1-\tfrac{r_0^2}{r^2}\big)} + r^2\,\Big(\extd\varphi - \frac{r_+ r_0}{r^2}\,\extd\tau\Big)^2
}{eq:vp10}
which depends on the mass parameter $r_+$ and the rotation parameter $r_0$.
The action \eqref{eq:3daction} evaluated on flat space cosmologies \eqref{eq:vp10} leads to the free energy [$T=r_+^2/(2\pi r_0)$, $\Om=r_+/r_0$]
\eq{
F(T,\,\Omega)=-\frac{r_+^2}{8G}=-\frac{\pi^2T^2}{2G\Omega^2}
}{eq:vp11}
in accordance with \cite{Bagchi:2013lma}. The entropy derived from the free energy \eqref{eq:vp11} is consistent with independent derivations of entropy, $S=2\pi r_0/(4G)$, using either the Bekenstein--Hawking relation $S_{\textrm{\tiny BH}}=A_h/(4G)$ or a generalization of the Cardy formula for Galilean (or ultrarelativistic) conformal algebras \cite{Barnich:2012xq,Bagchi:2012xr} (see \cite{Bagchi:2013xxx} for a recent generalization that takes into account logarithmic corrections). We stress that consistency with the first law of thermodynamics, $\extd F=-S\,\extd T-J\,\extd \Omega$ [where $J=-r_+r_0/(4G)$], is only achieved for the choice $\alpha=\tfrac12$ in \eqref{eq:vp6}. For later comparison we note that the mass (as defined through the canonical charges) is given by $M=-F=r_+^2/(8G)$.

Comparing the free energy \eqref{eq:vp11} of flat space cosmologies with the free energy of `hot flat space', $F=-1/(8G)$ it was shown in \cite{Bagchi:2013lma} that there is a phase transition between these two spacetimes, similar to the Hawking--Page phase transition \cite{Hawking:1982dh}, with a critical temperature $T_c=1/(2\pi r_0)$.

Note that the factor $\tfrac12$ in the boundary term in \eqref{eq:3daction} is crucial to obtain the correct normalization of the free energy. While the importance of this factor should be evident on general grounds and from our AdS discussion in section \ref{se:2.2}, we illustrate this point now by explicitly evaluating the thermodynamics of flat space cosmologies when considering instead the standard Einstein--Hilbert action with the usual GHY boundary term. Since the bulk action vanishes on-shell, the only change as compared to before is that the free energy gets multiplied by a factor of $2$ relative to the result \eqref{eq:vp11}.
\eq{
F_{\textrm{\tiny wrong}}(T,\,\Omega)=-\frac{\pi^2T^2}{G\Omega^2}
}{eq:vp37}
Assuming now the validity of the first law, we then obtain an entropy
\eq{
S_{\textrm{\tiny wrong}}(T,\,\Omega)=-\frac{\partial F_{\textrm{\tiny wrong}}}{\partial T}\Big|_{\Om=\rm const.} = \frac{2\pi^2T}{G\Omega^2} = \frac{\pi r_0}{G} = 2\,\frac{A_h}{4G} = 2 S_{\textrm{\tiny BH}}
}{eq:vp36}
that differs from the expected Bekenstein--Hawking entropy $S_{\textrm{\tiny BH}}$ by a factor of $2$.
If we instead insist on the validity of the Bekenstein--Hawking entropy (which is confirmed independently through a Galilean CFT calculation \cite{Barnich:2012xq,Bagchi:2012xr}) then we violate instead the first law, $\extd F_{\textrm{\tiny wrong}} \neq - S_{\textrm{\tiny BH}}\, \extd T - J\,\extd\Omega$. Thus, as in our AdS discussion, the existence of a well defined variational principle is crucial to reproduce the thermodynamics of flat space cosmologies.

\subsection{1-point functions}\label{se:3.3}

Now that we have established the existence of a well defined variational principle for the action \eqref{eq:3daction}, we compute the 1-point functions by turning on sources in the bulk. 
As usual in holographic correspondences, the sources correspond to non-normalizable modes that solve the linearised EOM \cite{Witten:1998qj}. We call any mode 'non-normalizable' whenever it violates our Euclidean flat space boundary conditions \eqref{eq:vp5}, \eqref{eq:bcdiagonal}.

In appendix \ref{app:A} we provide the most general solution $\lin$ to the linearised EOM in a suitable gauge and discuss the conditions for normalizability. We find
\begin{subequations}
\label{eq:vp26maintext} 
\begin{align}
 \lin_{\tau\tau} &= -\lin_{rr} = 2\xi^1 + {\cal O}(1/r) \\
 \lin_{\tau\phi} &= r^2\,\partial_\tau\xi^0_\varphi + {\cal O}(1)\\
 \lin_{\varphi\varphi} &= 2r^2\,\partial_\varphi\xi^0_\varphi  + {\cal O}(\tau r)\\
 \lin_{r\tau} &= \lin_{r\varphi} = 0\,.
\end{align}
\end{subequations}
In order to switch on suitable non-normalizable modes that can act as sources we switch on the function $\xi_\varphi^0$ in \eqref{eq:vp26maintext}.

Having done so, we consider now the first variation \eqref{eq:var1} with $\alpha=\tfrac12$ for the class of metrics
\begin{equation}
\begin{aligned}
  g_{rr} &= h_{rr}(\phi) + h_{rr}^{(1)}(\tau,\phi)/r + {\cal O}(1/r^2) & g_{r\tau} &= h_{r\tau}(\phi) / r + {\cal O}(1/r^2) \\
  g_{\tau\tau} &= h_{\tau\tau}(\phi) + h_{\tau\tau}^{(1)}(\tau,\phi)/r + {\cal O}(1/r^2) & g_{\phi\phi} &= r^2 + h_{\phi\phi}^{(1)}(\tau,\phi) r + {\cal O}(1) \\
  g_{r\phi} &= h_{r\phi}(\tau,\phi) + h_{r\phi}^{(1)}(\tau,\phi)/r + {\cal O}(1/r^2) & g_{\tau\phi} &= h_{\tau\phi}(\tau,\phi) + {\cal O}(1/r) \; ,
\end{aligned}
\end{equation}
with variations of the form [again $\delta (h_{\tau\tau} h_{rr})=0 $]
\begin{equation}
\begin{aligned}
  \delta g_{rr} &= \delta h_{rr}(\phi) + {\cal O}(1/r) &\qquad \delta g_{r\tau} &= {\cal O}(1/r) \\
  \delta g_{\tau\tau} &= \delta h_{\tau\tau}(\phi) + {\cal O}(1/r) &\qquad \delta g_{\phi\phi} &= \delta h_{\phi\phi}^{(0)}(\tau,\phi) r^2 + {\cal O}(r) \\
  \delta g_{r\phi} &= {\cal O}(1) & \qquad \delta g_{\tau\phi} &= \delta h_{\tau\phi}^{(0)}(\tau,\phi) r^2 + {\cal O}(1) \; .
\end{aligned}
\end{equation}
These variations by construction are compatible with \eqref{eq:vp26maintext}, appropriately generalized to allow also small gauge fluctuations that lead away from the gauge choice imposed in the appendix \ref{app:A}. The fluctuations with superscript $(0)$ correspond to non-normalizable contributions, i.e., sources.

In this case the identities \eqref{eq:vp15} together with on-shell relations give rise to
\begin{multline}
  \left.\delta\Gamma\right|_{\mathrm{EOM}} =
  \frac{1}{32\pi G} \int_{\partial M}\!\!\!\!\!\! \mathrm{d}^2x \, \Big[\big(2 r_c h_{r\phi} + {\cal O}(1) \big) \,\partial_\tau \delta h_{\tau\phi}^{(0)} + 
    h_{\tau\tau} \,\delta h_{\phi\phi}^{(0)} 
    + \Big(
      \frac{h_{\tau\tau}^{(1)} \partial_\tau h_{r\phi}}{h_{\tau\tau}}
      \\
      + h_{r\tau} \partial_\phi \mathrm{ln} h_{\tau\tau}
      - 2 \partial_\phi h_{r\tau}
      - 2 h_{\tau\phi}
      + h_{r\phi} \partial_\tau h_{\phi\phi}^{(1)}
    \Big) \,\delta h_{\tau\phi}^{(0)} \Big] + {\cal O}(1/r_c) \; . 
\label{eq:vp31}
\end{multline}
Requiring a finite response when the radial cut-off is sent to infinity, $r_c\to\infty$, restricts the sources to satisfy $\partial_\tau \delta h_{\tau\phi}^{(0)}=0$. This means the function $\xi_\varphi^{(0)}$ appearing in \eqref{eq:vp26maintext} can only be linear in $\tau$,
\eq{
\xi_\varphi^{(0)}=\xi_J(\varphi)\,\tau + \frac12\,\int_\varphi \!\extd\varphi^\prime\,\xi_M(\varphi^\prime)
}{eq:vp32}
and thus depends on two free functions of $\varphi$ that we call $\xi_{M,\,J}$. [Note that adding a constant to $\xi_\varphi^{(0)}$ does not change the sources since the quantity $\xi_\varphi^{(0)}$ appears only with first derivatives in \eqref{eq:vp26maintext}, so the arbitrary integration constant in the second term in \eqref{eq:vp32} is irrelevant.] The corresponding sources are then parametrized by these two free functions and generate two independent response functions: 
\eq{
\delta h_{\varphi\varphi}^{(0)} = \delta \xi_M +2\tau\,\delta\xi_J^\prime\qquad \qquad\delta h_{\tau\varphi}^{(0)} = \delta \xi_J
}{eq:vp28}
In terms of these two functions the variation \eqref{eq:vp31} (after taking the limit $r_c\to\infty$) reads
\begin{multline}
  \left.\delta\Gamma\right|_{\mathrm{EOM}} =
  \frac{1}{32\pi G} \int_{\partial M}\!\!\!\!\!\! \mathrm{d}^2x \, \Big[
    h_{\tau\tau} \,\delta\big(2\xi_J^\prime\,\tau+\xi_M\big)
    + \Big(
      \frac{h_{\tau\tau}^{(1)} \partial_\tau h_{r\phi}}{h_{\tau\tau}}
      \\
      + h_{r\tau} \partial_\phi \mathrm{ln} h_{\tau\tau}
      - 2 \partial_\phi h_{r\tau}
      - 2 h_{\tau\phi}
      + h_{r\phi} \partial_\tau h_{\phi\phi}^{(1)}
    \Big) \,\delta\xi_J \Big] \; . 
\label{eq:vp30}
\end{multline}

For reasons explained already, our focus is exclusively on zero-mode solutions. In that case we can drop all derivative terms in the response functions and the source functions $\xi_{L,\,M}$. Then the variation \eqref{eq:vp30} establishes our main result
\eq{
  \left.\delta\Gamma\right|_{\mathrm{EOM}} =
  \frac{1}{32\pi G} \int_{\partial M}\!\!\!\!\!\! \mathrm{d}^2x \, \big(
    h_{\tau\tau} \,\delta \xi_M
    - 2h_{\tau\phi}\,\delta\xi_J \big) \; . 
}{eq:vp29}
We give the two response functions appearing in the variation \eqref{eq:vp29} suggestive names
\eq{
M = \frac{h_{\tau\tau}}{8 G}\qquad J=\frac{h_{\tau\phi}}{4G}
}{eq:vp27}
in terms of which we can rewrite our main result \eqref{eq:vp29} as
\eq{
  \left.\delta\Gamma\right|_{\mathrm{EOM}} =
  \frac{1}{2}\, \int_{\partial M}\!\!\!\!\!\! \mathrm{d}^2x \, \Big(\frac{M}{2\pi} \,\delta \xi_M - \frac{J}{2\pi}\,\delta\xi_J \Big) \; . 
}{eq:vp64}
The overall factor $\tfrac12$ is the same as in the definition of the stress tensor \eqref{eq:BY}.
The factors $2\pi$ in the denominators are exactly as in \eqref{eq:BYresult}. The signs are adjusted suitably for Euclidean signature.

We evaluate now the response functions for flat space
\begin{equation}
  \mathrm{d}s^2 = \mathrm{d}\tau^2 + \mathrm{d}r^2 + r^2 \mathrm{d}\phi^2
\end{equation}
and find 
\eq{
M_{\textrm{\tiny flat}} = \frac{1}{8G} \qquad J_{\textrm{\tiny flat}} = 0\,.
}{eq:vp63}
The expressions \eqref{eq:vp63} coincide precisely with mass and angular momentum of flat space \cite{Barnich:2006av,Bagchi:2012yk}.

Similarly, we calculate the response functions for flat space cosmologies
\begin{equation}
  \mathrm{d}s^2 =
    r_+^2 \left( 1 - \frac{r_0^2}{r^2} \right) \mathrm{d}\tau^2 +
    \frac{\mathrm{d}r^2}{r_+^2 \left( 1 - \frac{r_0^2}{r^2} \right)} +
    r^2 \left( \mathrm{d}\phi - \frac{r_+ r_0}{r^2} \mathrm{d}\tau \right)^2
\end{equation}
and find
\eq{
M_{\textrm{\tiny FSC}} = \frac{r_+^2}{8G} \qquad J_{\textrm{\tiny FSC}} = -\frac{r_+r_0}{4G} \,.
}{eq:vp62}
Again the expressions \eqref{eq:vp62} coincide precisely with mass and angular momentum of flat space cosmologies \cite{Barnich:2006av,Bagchi:2012yk}, as well as with the thermodynamical expressions we derived in section \ref{se:3.2}.

In conclusion, we recover as 1-point functions for zero mode solutions precisely the response functions expected from previous canonical and thermodynamical analyses. This further confirms the validity of the action \eqref{eq:3daction} with one half of the usual GHY boundary term.

\section{Outlook}\label{se:4}

After reviewing how to obtain a well-defined variational principle for Einstein gravity with (Euclidean) AdS$_3$ boundary conditions and recovering known results for 0- and 1-point functions we applied the same methods to flat space boundary conditions. One of our main results is that the bulk action has to be supplemented by {\em one half} of the GHY boundary term. We exploited this result to recover thermodynamics of flat space cosmologies and to calculate the 1-point functions for zero mode solutions. We stress that the factor $\tfrac12$ in the boundary term was crucial to obtain the correct free energy \cite{Bagchi:2013lma}, since without this factor either the first law does not hold or the Bekenstein--Hawking relation is violated. Recent work that overlaps with ours \cite{Costa:2013vza,Fareghbal:2013ifa} does not use this boundary term but instead takes the AdS results for 1-point functions and performs a suitable limit of vanishing cosmological constant. The results of these papers for the response functions agree with our corresponding results, which were directly calculated in flat space. Thus, our results provide a proof of the smoothness of 0- and 1-point functions in the limit of AdS with vanishing cosmological constant.

A technical reason for our restriction to zero mode solutions was that the transformation between EF and diagonal gauge only works straightforwardly for such solutions. It would be interesting to generalize the discussion and to describe non-zero mode solutions in Euclidean signature. While our Euclidean boundary conditions allow also $\varphi$-dependent functions in \eqref{eq:vp5} consistent with a variational principle, the relation of such non-zero mode configurations to configurations in EF gauge is not clear to us. We leave this issue as an open problem, and just mention that its resolution probably requires to set up boundary conditions that are valid both at lightlike infinity and at spatial infinity. Physically, the restriction to zero mode solutions was sufficient for our purposes, namely to check the consistency of the free energy and the 1-point functions for flat space cosmologies.

\section*{Acknowledgments}

We are grateful to Arjun Bagchi for a pleasant and fruitful collaboration on related topics. We thank Tom\'as Andrade, Glenn Barnich, Geoffrey Comp\`ere, Reza Fareghbal, Don Marolf, Ali Naseh, Ioannis Papadimitriou, Alfredo Perez, David Tempo and Ricardo Troncoso for discussions. 

S.D. is a Research Associate of the Fonds de la Recherche Scientifique F.R.S.-FNRS (Belgium). DG and FS were supported by the START project Y~435-N16 of the Austrian Science Fund (FWF) and the FWF projects I~952-N16 and I~1030-N27. The work of JS was partially supported by the Engineering and Physical Sciences Research Council (EPSRC) [grant number EP/G007985/1] and the Science and Technology Facilities Council (STFC) [grant number ST/J000329/1].

\appendix
 
\section{Linearised solutions in flat space Einstein gravity}\label{app:A}

Any solution to the linearised Einstein equations in three dimensions must be locally pure gauge.
\begin{equation}
  \lin_{\mu\nu} = \nabla_\mu\xi_\nu + \nabla_\nu\xi_\mu = \partial_\mu \xi_\nu + \partial_\nu\xi_\mu - 2\Gamma_{\mu\nu}^\alpha \,\xi_\alpha
\label{eq:vp21}
\end{equation}
For a Euclidean background in polar coordinates the only non-trivial Christoffel symbols are $\Gamma^r{}_{\varphi\varphi}=-r$ and $\Gamma^\varphi{}_{r\varphi}=\Gamma^\varphi{}_{\varphi r}=1/r$.
Since transverse-traceless gauge does not lead to sensible results in flat space \cite{Bagchi:2012yk} we impose instead axial gauge on the fluctuations.
\begin{equation}
  \lin_{r\tau}=\lin_{r\varphi}=0\qquad \lin_{rr}=-\lin_{\tau\tau}
\label{eq:vp20}
\end{equation}
The last condition makes sure that the relation between $\lin_{rr}$ and $\lin_{\tau\tau}$ required by our boundary conditions \eqref{eq:vp8} remains intact.
Globally, linearised solutions of the form \eqref{eq:vp21} are not necessarily pure gauge, since they can have non-trivial canonical boundary charges. Fluctuations $\lin$ that are pure gauge even globally will be referred to as `small gauge transformations'.

To construct the most general solution to the linearised EOM, we are thus left to find the most general vector field $\xi$ preserving the axial gauge \eqref{eq:vp20}. This gives rise to three conditions that can be fully integrated. Preserving $\lin_{rr}+\lin_{\tau\tau}=0$ gives rise to
\begin{equation}
  \partial_r \xi_r + \partial_\tau\xi_\tau=0 \,.
 \label{eq:int1}
\end{equation}
Preserving $\lin_{r\tau}=0$ requires
\begin{equation}
  \partial_r\xi_\tau + \partial_\tau\xi_r = 0\,.
 \label{eq:int2}
\end{equation}
Both conditions together imply 
\begin{align}
 \xi_\tau &= \xi_\tau^0(\varphi)+\xi^+(r+\tau,\,\varphi) + \xi^-(r-\tau,\,\varphi) \\
 \xi_r &= \xi_r^0(\varphi)-\xi^+(r+\tau,\,\varphi) + \xi^-(r-\tau,\,\varphi)\;.
\end{align}
Finally, preservation of the gauge condition $\lin_{r\varphi}=0$ yields 
\begin{equation}
  \partial_r\xi_\varphi + \partial_\varphi\xi_r - \frac{2}{r}\xi_\varphi = 0\;.
 \label{eq:int3}
\end{equation}
This differential equation is solved by
\eq{
\xi_\varphi = r^2\,\xi_\varphi^0(\tau,\,\varphi) + \tilde\xi_\varphi\;,
}{eq:vp22}
where $\tilde\xi_\varphi$ is a particular solution of \eqref{eq:int3}, whose precise form depends on the functions $\xi^\pm$ and $\xi_r^0$.
The function $\xi_\varphi^0(\tau,\,\varphi)$ arises as a homogeneous solution and will play an important role for turning on sources.

Any solution for the vector field $\xi$ above then leads to a solution for the linearised fluctuations $\lin$:
\begin{subequations}
\label{eq:vp24} 
\begin{align}
 \lin_{\tau\tau} &= -\lin_{rr} = 2\partial_\tau\xi_\tau \\
 \lin_{\tau\phi} &= \partial_\tau\xi_\varphi + \partial_\varphi\xi_\tau \\
 \lin_{\varphi\varphi} &= 2\left(\partial_\varphi\xi_\varphi + r\,\xi_r\right) \\
 \lin_{r\tau} &= \lin_{r\varphi} = 0
\end{align}
\end{subequations}

Let us consider now solutions to the linearised EOM compatible with our boundary conditions \eqref{eq:bcdiagonal}. Then the functions $\xi^\pm$ above can be at most linear in $r\pm\tau$ for large values of $r$.
\begin{align}
 \xi_\tau &= \xi_\tau^0(\varphi)+r\,\xi^0(\varphi) + \tau\,\xi^1(\varphi) + {\cal O}(1/r) \\
 \xi_r &= \xi_r^0(\varphi)-r\,\xi^1(\varphi) - \tau\,\xi^0(\varphi) + {\cal O}(1/r)
\end{align}
Solving the differential equation \eqref{eq:int3} yields
\eq{
\xi_\varphi = r^2\,\xi_\varphi^0(\tau,\,\varphi) + r^2\,\ln r\,(\xi^1)^\prime - \tau r\,(\xi^0)^\prime + r\,(\xi^0_r)^\prime + {\cal O}(1)\;.
}{eq:vp23}
Compatibility with periodicity in $\varphi$ and the boundary conditions \eqref{eq:bcdiagonal} requires $\partial_\tau\xi_\varphi^0=0$, $\partial_\varphi\xi_\varphi^0=\xi^1$ and $(\xi^1)^\prime=0$, so with no loss of generality we set $\xi_\varphi^0=\phi\xi^1$ (we set to zero an additive constant to $\xi_\varphi^0$ since it does not contribute to the metric). There are no further conditions emerging from \eqref{eq:bcdiagonal}. 

Thus, the most general result for the vector field compatible with periodicity, gauge and boundary conditions of the metric is given by
\begin{align}
 \xi_\tau &= \xi_\tau^0(\varphi)+r\,\xi^0(\varphi) + \tau\,\xi^1 + {\cal O}(1/r) \\
 \xi_r &= \xi_r^0(\varphi) - r\,\xi^1 - \tau\,\xi^0(\varphi) + {\cal O}(1/r) \\
 \xi_\varphi &= \phi r^2\,\xi^1 - \tau r\,(\xi^0)^\prime + r\,(\xi^0_r)^\prime +{\cal O}(1)\;.
\end{align}
Inserting this result into \eqref{eq:vp24} establishes the normalizable solutions\footnote{As stressed in section \ref{se:3.3}, in this work we defined normalizable modes to mean solutions to the linearised EOM compatible with our boundary conditions.  Normalizability also has a definition in terms of the symplectic norm of the corresponding linearised solution, defined as \begin{equation}
  (\psi, \psi) = -i \int_{\Sigma_t} n^\alpha \omega_\alpha (\psi,\psi^*, \bar g), \nonumber
\end{equation}
where $\Sigma_t$ is a space-like surface (here $t=\rm const.$), $n^\alpha$ its normal, and $\omega_\alpha$ the symplectic current (see e.g. \cite{Andrade:2009ae}). From the asymptotic behaviour (\ref{eq:vp25}), one can check that the contribution to the norm from the $r=\infty$ boundary is finite (actually, vanishes), which is expected for linearized perturbations satisfying the boundary conditions \eqref{eq:vp8}. 
} to the linearised EOM: 
\begin{subequations}
\label{eq:vp25} 
\begin{align}
 \lin_{\tau\tau} &= -\lin_{rr} = 2\,\xi^1 + {\cal O}(1/r) \\
 \lin_{\tau\phi} &= {\cal O}(1)\\
 \lin_{\varphi\varphi} &= -2\tau r\,\big(\xi^0+(\xi^0)''\big) + 2r\,\big(\xi_r^0+(\xi_r^0)''\big)  + {\cal O}(1)\\
 \lin_{r\tau} &= \lin_{r\varphi} = 0
\end{align}
\end{subequations}

If we preserve the linearised on-shell property but violate our boundary conditions, we can generate non-normalizable modes. In particular, if we further turn on the homogeneous solution $\xi_\varphi^0(\tau,\,\varphi)$ in \eqref{eq:vp22}, vector fields of the form 
\begin{align}
 \xi_\tau &= {\cal O}(\textrm{normalizable}) \\
 \xi_r &= {\cal O}(\textrm{normalizable}) \\
 \xi_\varphi &= r^2\,\xi^0_\varphi  + {\cal O}(\textrm{normalizable})
\end{align}
lead to specific non-normalizable solutions:
\begin{subequations}
\label{eq:vp26} 
\begin{align}
 \lin_{\tau\tau} &= -\lin_{rr} = 2\xi^1 + {\cal O}(1/r) \\
 \lin_{\tau\phi} &= r^2\,\partial_\tau\xi^0_\varphi + {\cal O}(1)\\
 \lin_{\varphi\varphi} &= 2r^2\,\partial_\varphi\xi^0_\varphi  + {\cal O}(\tau r)\\
 \lin_{r\tau} &= \lin_{r\varphi} = 0
\end{align}
\end{subequations}
In section \ref{se:3.3} we show that the non-normalizable solutions \eqref{eq:vp26} appear as sources for the corresponding operators in the 1-point functions.

 
\providecommand{\href}[2]{#2}\begingroup\raggedright\endgroup

\end{document}